\documentclass[times, 12pt]{article}
\usepackage{epsfig}
\usepackage{mystyle}
\usepackage{law}
\usepackage{wrapfig}

\newlength{\lineheight}
\newcounter{saved}
\def\save{\saveitem{saved}\addtocounter{saved}{-1}}

\def\resume{\stepcounter{saved}\setitem{saved}}
{\begin{rules} \resume}%
{\save \end{rules}}

\title{Bracing Heterogeneous Distribted Systems\\
via Built-in Frameworks}
\author{Naftaly H. Minsky\\
Rutgers University \\
Department of Computer Science\\
Email: \TT{minsky@rutgers.edu} \\ }

\begin{document}

\maketitle
\bibliographystyle{plain}
\begin{abstract}
This paper introduces  a novel architecture  of distributed
systems---called  \emph{framed distributed system}, or FDS---that braces a given
system via a built-in \emph{virtual framework} that controls the flow of messages 
 between system  components, and between them and their
environment---while being oblivious of the
 code of the communicating components.
This control is carried out in a decentralized, and thus scalable, manner.

The FDS architecture is expected to  have a significant impact on the
dependability and security of distributed systems, and on  the whole life cycle
of
such systems.

Although this architecture has been designed specifically for  SOA-like
 heterogeneous and \emph{open} systems---whose components may be written in different
languages, may run on different platforms, and may be designed, constructed,
and even maintained under different administrative domains---it should be useful
for distributed systems in general.
\end{abstract}

\s{Introduction}\label{intro}
Heterogeneous distributed systems suffers from  serious difficulties in
establishing reliable
 \emph{system properties}---i.e., properties 
    that span an    entire system; or that involve a set of components
    dispersed throughout the system, such as a coordination protocol between
    the members of collaborating group of components.
The difficulties are in: (a) the reliable \emph{implementation} of such properties;
(b) the \emph{verification} that a property of this kind has been implemented
    correctly; and (c) the \emph{dependability} of system properties, namely,
 their  resilience to failures and attacks, and 
their stability   with respect to the evolution of the system.

Dealing with these difficulties is particularly hard
in systems whose components  may be written in different languages, may run on different platforms,
and may be designed, constructed, and even maintained under different
administrative  domains.   Such  systems are often said to have 
an   \emph{open architecture}, or just being
\emph{open}\footnote{The term ``open,'' as used here, has nothing to do 
 with the concept of \emph{open source}.}  \cite{kah88-1,bid98-1}---because of
 the  lack of effective constraints on the organization of the system as a whole,
 and on the internals of its disparate components.
  Systems are increasingly designed to be open, with the hope that
this would make them more flexible. 
 The concept of \emph{service oriented architecture}
\cite{pap06-1} (SOA)
is a prominent example of this trend, which is being adopted by a wide range of
complex distributed systems, such as: commercial enterprises, societal and 
 governmental institutions, and various types of \emph{virtual organizations} (VOs).
Moreover, even \emph{virtually monolithic} systems---namely, systems that are
constructed according to a single overall design,
 and maintained by a single organization---tend to become partially open
in time, because of the difficulties
 in controlling their evolution; and when several organizations, each with its
 own monolithic software, merge, forming some kind of federations.

To illustrate the problematic nature of    both the reliable implementation   of system
properties and their dependability  in open systems, consider the
following simple example: 
\begin{quote}
 Let $v$ be a widely used, and vital, component of an open distributed system,
 called Acme.
And suppose that for $v$  to be able to protect itself against  denial of service caused by too many
messages send to it from other components of Acme, this system is  designed to satisfy
the following (simplified) \emph{rate-control} (RC) protocol\footnote{This
protocol is oversimplified; a more proper protocol would buffer
messages sent too fast, and send them to $v$ automatically as fast as possible,
subject to the constraint imposed by $v$. \label{foot-a}}:

\begin{quote}
\emph{Once an arbitrary component $x$ of Acme gets a
message \texttt{slowDownTo(r)} from $v$, it would limit the rate of sending
messages to $v$ to the specified rate  \texttt{r}. }
\end{quote}
\end{quote}
\noindent The conventional, and seemingly natural, approach for establishing
such a global protocol  as a system-regularity---namely, ensuring that it is really
observed everywhere in the system---is \emph{code based}.  That is, 
 one should program each system component carefully to comply with the
given protocol.   But doing so  everywhere in a large and heterogeneous system,
is laborious,  error prone, and hard to verify.   And even if the RC protocol
 is established correctly in this manner, say by distributing  communication stabs that comply with RC
  to all system components---which is possible when all the components are
  written in a single language---it would not be easy to verify that these
  stabs are adopted correctly
 everywhere in the system. So, the code-based implementation is not very
 reliable. It is also not dependable, because it can be easily violated by an
 inadvertent or malicious change in any component. (Note that
the TCP flow-control might not help $v$, because the issue here is not
messaging congestion in the Internet around the sender of a message and its receiver, but the ability of $v$ to handle
messages.)

In a local (i.e., not distributed) system, written in a suitable language, one
may implement the RC protocol  via techniques like   \emph{reflection},
 as under  \emph{meta object protocol}
(MOP) \cite{kic91-1}; or  via  \emph{aspect oriented
programming} (AOP) \cite{kic01-1}.
AOP can be used even 
for  a centrally managed distributed system, if all its components are written in a
single language that support AOP; but it cannot be used for a
 truly open system.

The approach advocated by the proponents of SOA for addressing
such problems involves ``good software engineering practices,'' and
what is called 
``governance techniques''~\cite{pap06-1}---where ``governance'' means managerial
 techniques for establishing  policies about the processes of
 creation and evolution of the system at hand. Such a human-based
  discipline is indeed necessary, but it is not sufficient;   there is a  need
  for  a more rigorous, and  more effective approach for addressing these  fundamental difficulties
  with open distributed systems.

\p{The Contribution of this Paper:}
We address here these basic difficulties by introducing the concept
of \emph{framed distributed system} (FDS), which is a system that
 operates  and evolves  under  a
 \emph{framework} that governs  the  exchange of messages 
 between system  components, and between them and their
environment. More specifically,
the framework of an FDS   consists of  a set of  strictly enforced \emph{laws} about
the flow of messages in the system;
laws that are
 \emph{oblivious of the
 code of the components that send and receive such messages}.
Such obliviousness is necessary
 for open systems, because of the lack of global knowledge of, and control over, the
 code of many of the their  components.  And it also 
has the advantage of rendering the framework, and the properties established by
it,  independent of the code that populate the various
system components, and thus invariant of the evolution of this code.

Of course, the Independence  of the code of system components also means that the
framework  can have only marginal effect on the
functionality of the system. But it is our thesis that a framework can have
significant impact on important non-functional qualities of the system governed
by it, such as enhancing its dependability.

 The framework of an FDS is different
from the concept of \emph{architectural model} \cite{gar95-1,med99-1}, which is an
external specification of certain aspect of a system that are required to be
implemented by its code. This  model is generally
not enforced, it is therefor prone to a gap between the model and the actual system
behavior, as is well known \cite{mur95-1}.
However, since the framework of an FDS is enforced,  the framework
can be considered as an integral part of the system---although it is not part
of any system component. 
 One can draw an analogy between the  framework of an FDS and
  the rigid metal
 framework of a building. The latter has little effect on  the internals of the
 building such as its internal walls,   doors, and  windows; and  on  how the
 building is to function. And yet, this metal framework
 provides its building 
 with an indispensable degree of dependability, stability and safely.
The framework of an FDS is expected to provide analogous benefits.

\p{The Structure of this  Paper:} 
The rest of this paper is organized as follows.
\secRef{principles} introduces the principles on which the concept of FDS is based.
 \secRef{lgi}  describes briefly the middleware---called Law Governed
 Interaction (LGI)---employed by
 our FDS architecture.
\secRef{model} introduces the architecture of framed distributed systems.
\secRef{case} describes an implemented case study of an FDS,
 which provides a concrete realization of this architecture.
  \secRef{impact} discusses the potential impact of FDS on the
 dependability of distributed systems, focusing on fault tolerance at the
 application level of systems; (we expect the FDS architecture to have a broad impact on
the entire life cycle of distributed systems: on their design,
 construction, and evolution---but a detailed discussion of such impact is
 beyond the scope of this paper).
  \secRef{related}  discusses work  related to our concept of FDS, such as research on
\emph{policy based frameworks}.
\secRef{future} discusses some open problems raised by the FDS architecture.
And   \secRef{conclusion} concludes this paper.

\textbf{A comment about terminology:}
 We will replace the  common term
``component,'' used above,  with the term ``actor''---which
may be any  autonomous process of computation that sends and receives messages. The term
actor is meant to reflect
the fact that  the framework of an FDS is oblivious of the internals of
the components of the distributed system governed by it, focusing only on their externally observable activities, namely
on the exchange of messages between them. We will, nevertheless,  use occasionally the
term ``component,'' particularly when discussing related work.

\s{The Principles Underlying the FDS Architecture}\label{principles}
We spell  out here the underlying principle of the  FDS architecture, which
are  the key properties  that  an FDS is
 to satisfy---along with their rationale.

\begin{enumerate}
\item \emph{While the framework of an FDS is to regulate the flow of messages in a system, it is
to be oblivious of the
 code of the components that send and receive such messages}. 
We have pointed out in the Introduction that obliviousness of the code is
necessary for open systems, because of the lack of global knowledge of, and
control over much of this code.
 Moreover, this principle has several advantages, even for monolithic systems,
 as follows: (a) it makes the framework independent of the languages in which
 the various system components are written; (b) it simplifies the enforcement of
 laws, by not requiring the laws of the framework to be \emph{weaved} into the
 code of components, as under AOP \cite{kic01-1}; and (c) it enhances the dependability  of
the  constraints imposed by the framework, in that they are
 invariant  of the evolution of the code.

\item \emph{The framework should be able to be   sensitive to the history of
    interaction---i.e., it should be stateful.} 
This is required  to enable the imposition of   inherently stateful
    coordination protocols, such as the rate-control  protocol discussed above,
 or the \emph{choreographies} under SOA \cite{par09-1}.
\item \emph{The enforcement of the laws of a framework should to be decentralized.}
This is required for several reasons: 
first, for scalability, because the enforcement of stateful constraint via a
    single, even if replicated, reference monitor is unscalable;
second, for preventing  the enforcement mechanism
itself from becoming  a
single point of failure; and third, for avoiding having   a central target for  attacks.

\item \emph{Law enforcement must be strict.}  By ``strict enforcement'' we
      mean, for example, that messages  prohibited by the framework
 should not be  transfered.
  We note that strict enforcement is necessary for dependability---one cannot
  depend on a law that may or may not be followed by the system\footnote{The
  strict
 enforcement is the main reason for using the term ``law'' for what is often called 
 ``policy,'' particularly in the 
access control (AC) literature. A policy generally means a constraint, or a
 plan of action, that may
or may not be enforced. We are using the term law in some analogy to
 the \emph{laws of nature}, which are rules that are known to be satisfied.}.
We not however, that 
 that laws themselves may be  \emph{lenient} in various respects. In particular, a law
 may explicitly allow an
 undesirable message to be transferred as is, requiring  
a copy of this message to be sent to some monitor or manager,
 for future disposition.

\item \emph{A degree of trust needs to be established between all actors of a
      given FDS, regarding their interactive behavior:}
Such trust is obviously necessary  for effective interaction between the
various actors of a system, and there is normally a very little basis for such
trust in open systems, such as under SOA.

\item \emph{The structure of the framework should be highly modular.}
Modularity is required because
the body of laws that needs to be imposed over the
flow of messages in 
 a large and heterogeneous distributed system is unlikely to  be monolithic.
 It would generally be  composed of multitude of
 diverse and semi independent laws, which may be formulated
by different stakeholders, at different times, with little or no coordination
with each other.
 One  such  law
 may impose global constraints over the entire system in question.
Other laws may govern
different parts of the system---e.g., different divisions
that may belong to different administrative domains.
Still other types of laws may govern groups of actors dispersed throughout the
system, which are 
 involved in some collaborative activity, imposing a suitable coordination
 protocol on them.  And many of these laws may have to \emph{conform} to
 others---in particular, all of them would need to conform to the global system
 law. 

Such a collection of distinct laws need to be represented as modules of the framework,  in a
manner that would facilitates their incremental  construction and evolution, 
 by different stake-holder. We will satisfy these requirements via what we will
 call \emph{conformance hierarchy of laws}.

\item \emph{The evolution of the framework of an FDS should be controllable:} 
This is important because a change of the framework might have a strong effect
on the system it governs. 
And the simplest way to enable such control is to make the framework self-regulatory.
\end{enumerate}
\noindent
We shall see in due course how these principles are satisfied by the
FDS architecture.

\s{The (LGI) Middleware---a Partial Overview}\label{lgi}
The FDS architecture requires a suitable middleware for supporting the
principles stated in \secRef{principles}. We
have chosen for this purpose a middleware called \emph{law governed
interaction} (LGI),  which had been developed  by the author and his
students. LGI is broadly related to 
the  access control (AC) mechanisms such as RBAC~\cite{osb00-1} and XACML~\cite{god05-1}, that regulate the
exchange of messages between the members of a collection of distributed actors.
But LGI differs from AC in  several
 fundamental ways, three of which are the following:
(1) AC has been designed to permit access---via messaging---only to those that have the right for it, and  without much concern
about the  the dynamic nature of message exchange.
LGI, on the other hand, can be fully sensitive to the history of interactions.
(2) LGI replaces the virtually centralized enforcement of its policies with
a decentralized enforcement, which is scalable even for stateful laws.  (3) LGI replaces the concept of \emph{policy} used by AC, with a very
    different, and considerably more general, concept of \emph{law}---which, in
    particular, unifies the concepts of mandatory and discretionary policies,
    views as distinct under AC. And (4)  the structure of LGI-laws facilitates
    their organization into a modular
    \emph{conformance hierarchy}, which is critical to FDS, and which has no
    parallel under AC.
Some of these properties will be discussed briefly below; for a full discussion
of  these, and other, differences between AC and LGI, see
\cite{min03-6,min12-2}. 

Here we present a partial overview of LGI,  focusing on 
 the following three key aspects of it, which are most relevant to this paper:
(1) the local nature of LGI laws; (2) their decentralized enforcement; and  (3)
    the handling  of multitude of interrelated laws.
  A more detailed
presentation of this middleware, and a tutorial of it, can be found in its manual
\cite{min05-8}---which describes the release of an experimental implementation
of the main parts of LGI.  For additional
information  the reader is referred to a host of published
papers, some of which will be cited  in due course.

\ss{LGI Laws, and their Local Nature}\label{local}
 Although the purpose of LGI is to govern the exchange of messages between
 different distributed actors,
the LGI laws do not do so directly. Rather, a law governs the \emph{interactive
activities} of any actor operating under it, in particular,  by imposing
constraints on the messages that such an actor can send and receive.

 A  \emph{law} \EL\ is defined over three
elements---described with respect to a given  actor $x$  that operates under
this law:  (1) A set $E$ of \emph{interactive events} 
that may occur at any actor, including the arrival of a message at $x$, and
the sending of a message by it.
(2)  The  \emph{control-state} (or, simply, state) $S_x$  associated
 with $x$---which  is distinct from the internal state of $x$, of which the
 law is oblivious. 
 And (3) a set $O$ of
 \emph{interactive operations}---such as forwarding a message and 
accepting one---that can
 be mandated by a law, to be carried out at $x$ upon the occurrence
 of interactive events at it.

Now, the role of a law  is to decide what should be done in response to
the occurrence of any interactive event at an actor  operating under it.
This decision, with respect to an actor $x$,
is formally defined by the following mapping:
\begin{equation}
 E \times S_x \rightarrow  S_x \times (O)^*.
\label{eq-law-2}
\end{equation}
In other words,  for any  a given  $(event,state)$ pair, the law mandates a new
state, 
as well as a (possibly empty) sequence of  interactive operations to be carried
out at $x$.
 Note, in particular, that
the ruling of the law upon the occurrence of an  event depends on the state of $x$ at
that moment; and that the same law determines how the state can change. 
 LGI laws are, therefore, \emph{stateful}---i.e.,  \emph{sensitive
to the history of the interactive-events}, at a given  actor $x$.
Moreover, although this is not evident from the above abstract definition,
an LGI law can be
 \emph{proactive}, 
in that it can force some messages to emanate from an actor, under  certain
circumstances, even if the actors itself did not send such messages---thus
these  laws can ensure  both \emph{safety and liveness} properties.

Note that LGI laws are \emph{local} in the sense that they depends only the
occurrence of events at a single actor, and on the interactive state of this
actor alone; and a law can effect directly only the interactive behavior of the
actor operating under it.
 It is worth pointing out that   although locality constitutes a strict constraint on the structure
of LGI laws, it does not reduce their expressive power, as has been proved in
\cite{min05-8}. 
In particular, despite its \emph{structural locality}, an LGI law can have
global sway over a set of actors operating under it.

Finally, note that the law is a complete function, so  that any mapping of the type
defined above is considered a valid
law. This   means that a law of this form is \emph{inherently self
consistent}---although a law can,  of course,  be wrong in the sense that
it may not work as
intended by its designer.

\p{About Languages for Writing Laws:}  Formula~\ref{eq-law-2} is an abstract
definition of the semantics of laws. It does not, in particular,
 specify a language for
writing laws.
 In fact, the current implementation of LGI supports two different
\emph{law-languages}, one based on  Prolog, and
the other  on Java; and another simpler law-language is under development.
 But the choice of language has no effect on the
semantics of LGI, as long as the chosen language is sufficiently powerful to
specify all possible mappings defined by Formula~\ref{eq-law-2}.

Space limitation preclude the description of any of these languages, but to
give a sense of them we replicate in \figRef{fig:frequency} a law \law{CC} (for ``Congestion
Control'') written in the Prolog based law-language of LGI, which
essentially\footnote{This law differs from our RC protocol in several ways:
(a) it enables every actor (not just a single actor as $v$ under RC) to control the rate of messages sent to it (which is
    done via a message ``changeDelay'' instead of the ``slowDownTo'' under RC); (b) and its
    state initialization is done differently under present version of LGI. } 
represents the RC protocol introduced in \secRef{intro}. This law is explained
in detail in \cite{min97-1}, where it was first introduced.

\begin{ruleset}{Law  \CAL{L}$_{\cal CC}$ that establishes congestion control
\label{fig:frequency}}
\begin{small}
\init Each client has in its control state: (1) the term \TT{clock(T)} ,
where \TT{T} represents the local current time; (2) a term
\TT{delay(DT)} where \TT{DT} represents the minimum delay between
successive messages sent by the client to the server \TT{s}; and (3) a
term \TT{lastCall(Tlast)} where \TT{Tlast} is the time when the last
message was sent to the server (initially set to 0).

\Rule sent(s,_,_):- do(forward). \vspace{2mm}
\label{sentServerMessage}\vspace{4mm}

\Rule arrived(s,changeDelay(Val),X):-\\
	do(delay(DT) $\leftarrow $ delay(Val)),\\
        do(deliver(memo(changeDelay(Val)))).\vspace{2mm}
\label{arrivedChangeDelay}

\Rule arrived(_,_,_):- do(deliver). \vspace{2mm} 
\label{arrivedMessage}

\Rule sent(X,M,s) :-\\
	lastCall(Tlast)@CS,delay(DT)@CS,clock(T)@CS,\\
	T > (Tlast + DT),\\
	do(lastCall(Tlast) $\leftarrow $ lastCall(T)),do(forward).
\label{sentMessage}}
\end{small}
\end{ruleset}

\ss{The Decentralized Law Enforcement,\\ and the Concept of
\EL-\emph{agent}}\label{enforcement}
The local nature of laws enable their decentralized enforcement, because a law
 can be enforced
on every actor subject to it with no knowledge of, or dependency on, the simultaneous   interactive state of any
 other actor of the system.
Such enforcement is scalable even for highly stateful policies that are
sensitive to the history of interaction  (cf. \cite{min12-2}).
Here is how the enforcement of LGI works.

To  communicate under  a given LGI law \EL, an actor $x$ needs to engage
a generic software entity called \emph{controller}\footnote{Controllers are
actually hosted by \emph{controller-pools}, each of which can host a number of \emph{private
controllers}, which may operate under different laws.}, which generally does not
reside on the host of its patron $x$. The controller is
 built to mediate the interactive activities of any actor that engages  it,
 under any well formed law that the actor chooses.
Once such a controller  $T$ is engaged by an actor $x$, subject to a law \EL,  it becomes the
private mediator for the interactive activities of $x$, and
is   denoted by $T^{\mathcal{L}}_{x}$. 
  The   pair $\langle x, T^{\mathcal{L}}_{x}\rangle$
  is called an
\EL-\emph{agent}---or, more generally an \emph{LGI-agent}, and sometimes simply an \emph{agent}.  And a set of interacting
\EL-\emph{agent}, for a given law \EL, is called an \EL-\emph{community}.

\begin{figure}
\leavevmode
\epsfysize=1.2 in
\centerline{\epsffile{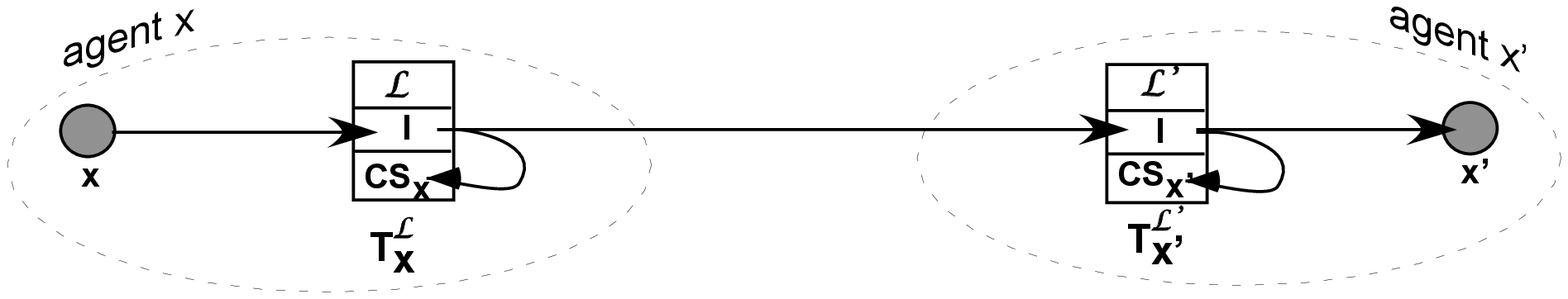}}
\caption{Interaction between a pair of  LGI-agents, mediated by a pair of
controllers under possibly different laws.
}
\label{fig-agent}
\end{figure}

 \figRef{fig-agent} depict the manner in which a pair of agents, operating
 under possibly different laws, exchange a message. (An agent is depicted
 here by a dashed oval that includes an  actor and its  controller.)
Note the  \emph{dual nature} of control exhibited here:
The transfer of a message is first mediated by the sender's controller, subject
to the sender's law, and then by the controller of the receiver, subject to its
law. This dual control, which is a direct consequence of the local nature of
LGI laws,
 has some important
consequences. In particular, it facilitates flexible interoperation, as
discussed in \secRef{interop}.

\p{Mutual Recognition:} It should be pointed out that
 a pair of interacting  LGI-agents can recognize each other as such, and can
 identify each other law by its one-way hash. This enables them to recognize
 when they operate under the same law, thus belonging to the same
 \EL-community.
And if they operate under different laws, they are able to get the 
 text of each other's law.

\p{About the Trustworthiness of Controllers:} Consider a set $S$ of agents interacting via LGI, and let $T_S$ be the set
of controllers employed by them.
$T_S$ is, essentially the  \emph{trusted computing base} (TCB) of $S$. 
There are several reasons for trusting the controllers in $T_S$, despite the fact that unlike most
TCBs,  $T_S$ is to be distributed. Some of these reasons are, briefly, as follows.

First, $T_S$ can be  maintained by what is called  a \emph{controller service}
(CoS), which is to be  managed by some trustworthy company---which may well be
the company, or the virtual organization that uses the CoS as its TCB.
Second, controllers are generic and, like language compilers, can be well
tested, and  thus
 more trustworthy than the disparate actors that use them.
Third, the distributed  $T_S$ is more fault tolerant than a single, central,
reference monitor, 
because it does not constitute a single point of failure. And, fourth, $T_S$ is
more secure than a central,
reference monitor, because it does not constitute a single point of attack.

\p{About Performance:} 
The overhead incurred by the LGI control turns out to be relatively small.
In circa 2000 it was measured to be around 50 microseconds for fairly common
laws, which is  negligible for
communication over WAN. This is one of the results of  a comprehensive study of
this overhead in \cite{min99-5}.

\ss{The Organization of Laws into a Conformance  Hierarchy}\label{conformance}
As pointed out in \secRef{principles}, a complex system may need to be governed by a
set of semi-independent laws. LGI enables the organization of such a
collection of laws, which  collectively governs a
single system, into  what is called a \emph{conformance hierarchy}. This is a tree of laws  rooted by law called \law{R}, in
 which every law, except of  \law{R} itself,  conforms transitively to its
 superior law, in a sense to be described below. Moreover the conformance
 relation between laws is inherent in the hierarchy, requiring no extra validation.
For a formal definition of such hierarchy of laws, and a detailed example of
its use, see \cite{min03-6}; here we provide just an informal introduction of
this concept.

\p{The Nature of  Conformance of LGI-Laws:}
Several access control  mechanisms  \cite{bel02-1,god05-1} defined conformance between policies basically as
follows: \emph{policy $P'$ conforms to policy $P$ if and only if $P'$ is  more
restrictive than  $P$, or   equal to it}. But this would not do for LGI-laws,
for several reasons, the most important of which is the following. The ruling of
an LGI-law is not confined to a decision whether to approve or reject an
action by an actor; it can also require some other actions to be carried out in
response to an event,
such as changing the state in a specified manner, or adding something to a
message being sent. And it is generally not meaningful to ask if one such
action is more or less restrictive than another.
So, instead of using a 
 uniform definition of conformance, based on restrictiveness,  LGI lets each law
define what it means for its subordinates to conform to it.
This is done, broadly, as follows.

 A law that  belongs to a conformance  hierarchy has two parts,
 called the \emph{ground} part and the \emph{meta} part.
The ground part of a law \EL\ imposes constraints on 
 interactive behavior of the
actors operating directly under this law---it
 has the structure defined by Formula~\ref{eq-law-2}. While the meta part of \EL\ 
 circumscribes  the extent to which   laws subordinate to \EL\ are allowed to
 deviate from its ground and meta parts.
In particular, this allows  a law, anywhere in this
hierarchy, to make any of its provisions \emph{irreversible} by any of its
subordinate law, by not permitting any deviation from it, by any of its
subordinate laws.

One application of such conformance is setting out defaults.
For  example, the
root law \law{R} may prohibit all interaction between components, while
enabling subordinate laws to permit such interaction, perhaps under certain
conditions.  Alternatively, law \law{R} may permit all interaction, while enabling
subordinate laws to prohibit selected interactions. 

This very flexible concept of conformance is somewhat analogous to the manner
in which the federal law of the US circumscribes the freedom of state laws to
deviate from it.
 Such conformance turns out to be also useful for the
governance of complex distributed systems, as we shall illustrate in
\secRef{case}.

\p{The Formation of a  Conformance Hierarchy of Laws:} 
A conformance hierarchy\footnote{We denote here a law-hierarchy by the symbol
$F$, because we will use $F$  to denote the framework of an FDS, which is such a hierarchy.}
  $F$ is formed incrementally via a   recursive process described
informally below.
First one creates the root law \law{R} of $F$.
Second,  given a law \EL\ already in $F$, one
defines a law  \CAL{L'}, subordinate to \EL, by means of 
 a law-like text called  \emph{delta}, denoted by
$\Delta($\CAL{L},\CAL{L'}$)$, which specifies the intended differences between
 \CAL{L'} and \EL.
Now,  law 
 \CAL{L'} is  derived dynamically from  law
\EL\ and $\Delta($\CAL{L},\CAL{L'}$)$,  essentially \emph{by dynamic consultation}, as
described informally below.

Consider the special case involving the 
root law \law{R}, and its subordinate law  \law{s} derived from \law{R} by the
 delta   $\Delta($\law{R},\law{s}$)$. And let  agent $x$ operate under law
 \law{s}. Now, 
when an event $e$ occurs at an agent $x$ it is first submitted to law \law{R} for evaluation. Law \law{R}
may consult the delta  $\Delta($\law{R},\law{s}$)$ of \law{s}
 before deciding on its
ruling---although it may also render its own ruling, not involving the delta.
If consulted,  the  delta will do its own evaluation of this event,  and will return its
\emph{advice} about the ruling to law \law{R}.
 \law{R} would render its final ruling about how to respond to event $e$,
  taking the
 advice of the delta into account---but not necessarily accepting it, because
 this advice  might contradict the meta part of \law{R}.
 In this way, the dynamically derived law \law{s} naturally conforms to its
superior law \law{R}, requiring no further verification.

A notable property of the hierarchical organization of laws is that
interacting agents operating under laws in a common hierarchy
 can identify the   position of each other's laws within this hierarchy.

\s{The Architecture of  Framed Distributed Systems}\label{model}
We start  by  introducing the  anatomy of a framed distributed system  (FDS), and then continue by
addressing the following aspects of such systems:
 (1) the construction of an FDS; (2) the trust modality induced by this
    architecture; (3) interoperation between actors operating under different laws; (4) the phenomenon of rogue
    communication, and its limited effect on an FDS;  and (5) the self
    regulatory nature of the framework of an FDS.

\ss{The Anatomy of an FDS}\label{anatomy}
 A \emph{framed distributed system}---or  an FDS---is
defined    as a triple $\langle F, A, C\rangle,$ where
 $F$ is the \emph{framework}, defined as a \emph{conformance hierarchy of
  laws} (a concept described in \secRef{conformance}), which is sometimes
referred  to as the  \emph{law ensemble} of the FDS;
$A$ is the \emph{set of  actors}, each of which exchanges messages
subject to some law in $F$ ($A$ is sometimes referred to as the \emph{base
system} of the FDS;
and  $C$ is the set of  generic
LGI  controllers that
 mediate the interactive activities of the actors of
$A$. We elaborate below on these elements of an FDS, and on the relationship
between them.

\begin{figure}
\leavevmode
\epsfysize=1.5 in
\epsfxsize=6.0 in
\centerline{\epsffile{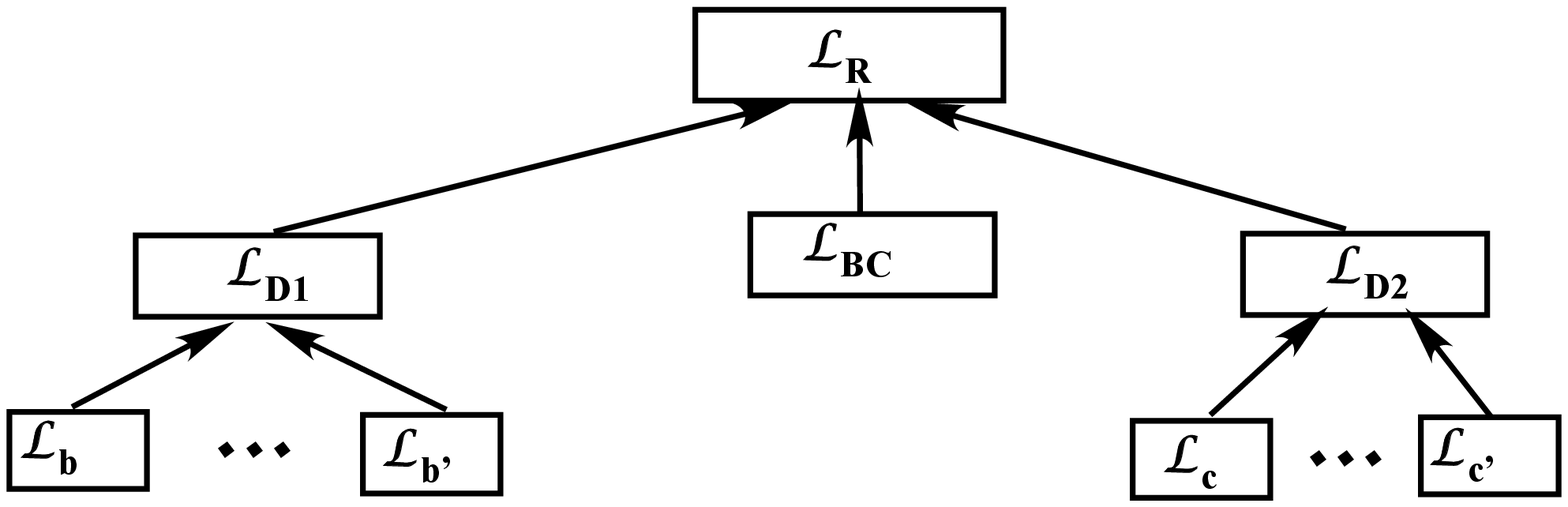}}
\caption{A Basic Hierarchical Law-Ensemble for the Acme System
\label{fig-laws}}
\end{figure}

 All the laws constituting the framework
$F$ are   maintained by a single  \emph{law-server}, denoted by
$LS$---which is, itself, a 
 member of $A$ (the significance of this fact is explained in \secRef{self-reg}). 
Note that the laws in $F$ make no assumptions about  the 
internals  of the  actors in $A$, which are viewed as black boxes by the
framework.
The structure of $F$  is exemplified by 
\figRef{fig-laws} that depicts  the framework
used by  the case study in
\secRef{case}. This particular ensemble of laws is a three level conformance
hierarchy, but it can, in general, be of any depth.

It should be pointed out that due to the conformance nature of the hierarchical
law ensemble $F$, its root law \law{R} has dominion over all the laws in it.
This dominion is absolute for any provision of \law{R} defined as
irreversible.
Other provisions of \law{R} may be  modified by subordinate laws, subject to
constraints imposed by \law{R} on their modification.
Consequently, this law governs, directly or indirectly, the entire framework $F$.

The function of the set $C$ of controllers---maintained by
some \emph{controller service} (CoS)---is to mediate the interactive
activities of actors in  the set $A$,  subject to various laws in $F$. 
Therefore,  this CoS  constitutes the \emph{trusted computing
base} (TCB) of the FDS---trusted to enforce the  laws of the framework $F$.
For an actor $x$ to operate subject to law \EL\ in $F$, it has to 
acquire a controller from the CoS, and engage it to operate under a law \EL\ in
$F$, thus
forming an \EL-agent, namely the pair
$\langle x,$ \Tla{L}{a}$\rangle$ 
of an actor with its controller. We will often refer to the agents thus operating
subject to laws in $F$, as $F$-agents.

Note that a single actor can animate several different $F$-agents, via
different controllers, operating subject to the same or different laws. This may be the
case, for example, when a single server provides several different services,
possibly subject to different laws.
Therefore, \emph{it is the $F$-agents that are the loci of control by the framework,
not the actors.} Not also that an actor $x$ that animates one or several $F$-agents
 of a given system $S$, may, at the same time, operate as part of other
systems, which may or may not be framed---this is particularly true under SOA,
whose components may be independent services,  which serve several systems.

\begin{figure}
\leavevmode
\epsfysize=3.3 in
\epsfxsize=4 in
\centerline{\epsffile{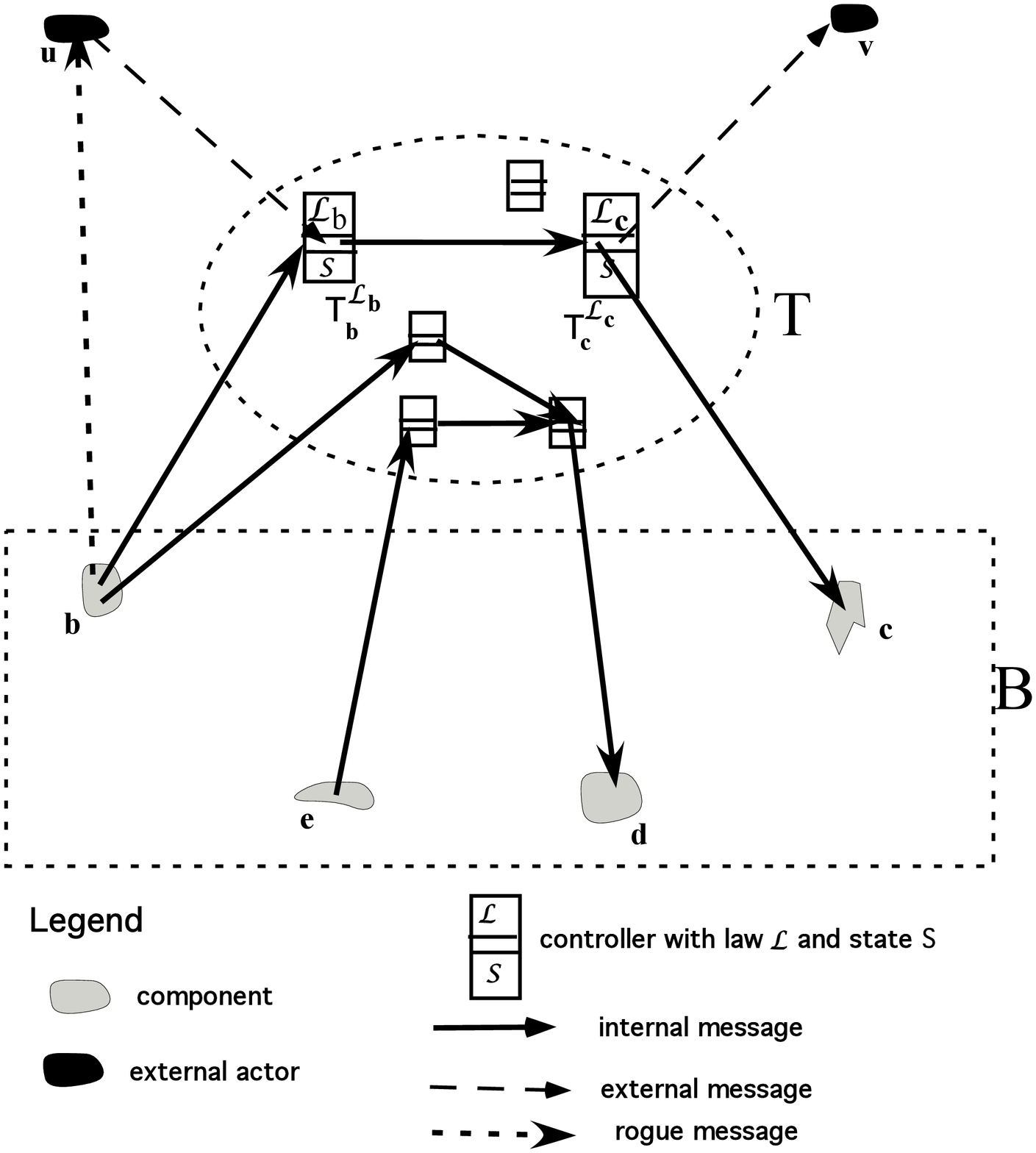}}
\caption{A Schematic Depiction of an FDS
\label{fig-arch}}
\end{figure}

An  FDS is portrayed schematically by
  \figRef{fig-arch}.
The  actors in $A$ are depicted within the dotted rectangle by  irregular
shaded figures,  representing
the presumed heterogeneity of such actors.
 The controllers belonging to $C$ 
are depicted by rectangles.
Finally,   the dark irregular shapes on top
 of this figure   depicts  actors that do not operate under laws in $F$,
 or under LGI at all---and thus do not belong to the system in question,
 although they
may  interact with its  $F$-agents, depending on the laws under which
these $F$-agents operate.
Note also  that  actor \textbf{b} 
 is depicted as animating two $F$-agents  operating via two
different controllers.

\ss{The Construction of an FDS}\label{deploy}
We describe here the  construction
of an FDS from scratch, which is quite straightforward.
A related, but more complex, issue is that of  the conversion of a
legacy  system  into an FDS,  which is still under investigation.
The construction of  a brand new FDS   can be described as consisting
of two consecutive phases: 
(1) the design and construction of the \emph{foundation} of  $S$; and (2)  the
    incremental  construction of the rest of it.

 The \emph{foundation} of an FDS $S$ consists of  two distinct part: (a) the
root law \law{R} of the  framework $F$ of $S$, which 
 can be viewed as part of the design of $S$, and would determine its overall
structure;
and (b) the  components, required for the definition of \law{R}.
These components must include the law-server $LS$ that maintains $F$,
and it may include some other components, such as a certification authority (CA) which law \law{R} may
 employ for the authentications of actors.

Once the foundation of $S$ is in place, the rest
 of it can be constructed \emph{incrementally}, via two kinds of steps:
(1) Adding a law  to $F$, subordinate to an existing law in it. And  (2)
    introducing a new $F$-agent into system $S$, by having an actor $x$ adopt
    some law \EL\ of $F$. This actor $x$ may already animate some $F$-agents of
    $S$; or it may be a new actor that have been built specifically for system
    $S$, and it can even be an autonomous actor that operates in several
    systems, as pointed out above.
But, as  we shall see in \secRef{case},
   a law may impose conditions on the actors attempting to adopt it, in
   particular,
by requiring actors to authenticate
themselves in a specified manner  in order to  operate under a given law.
And note again, as have already pointed out, that a given actor can form
several different $F$-agents, operating under possibly different laws.

These two types of additions to an FDS can be carried
    out by different stakeholder, and it can be done in various orders; and
    many of them can be done concurrently. Such incremental construction has
    two additional aspects that are discussed below.

\p{(1) Immunity from Inconsistencies}\label{immune}
A complex system, regulated by multiple policies, may, in general, suffer from
inconsistencies, particularly if these policies are formulated by different
stakes holder. Such inconsistencies   plagues many AC mechanisms, such as a set
of firewalls protecting an enterprise \cite{ioa00-1}, and the
XACML~\cite{god05-1} mechanism---both of which require techniques for resolving
such inconsistencies.

But the framework of an FDS is inherently free of inconsistencies.
 Indeed, 
 a single, monolithic law,  defined by  Formula~\ref{eq-law-2}, is self
consistent, as pointed out in \secRef{local}.  A single law in a hierarchy, which
is the result of applying a sequence of deltas to the root law is self
consistent by its construction, as described in \secRef{conformance}. 
Finally, 
 the set of laws constituting
a framework $F$ is  immune from inconsistencies with each other,
because  different laws govern the interactive activities of different agents,
so they cannot, by definition, be inconsistent---although they may refuse
to interoperate (c.f.   \secRef{interop}).

This inherent lack of inconsistencies
 facilitates the  construction and evolution of the framework of an FDS,
 and makes it
easier to reason about it. But of course,  being consistent does not mean
that a single law in $F$, or $F$ as a whole, cannot be wrong.  It is wrong
if it does not satisfy the intention of its  designers---which can, of course,
happen.

\p{(2) Laws do not Need to be  Diffused Among Controllers:}
Given the multiplicity of controllers, operating under different laws, one may
think that there is a need to diffuse carefully the right laws in the right
controllers. This is, essentially, what is done for 
enterprise systems  protected via a distributed set firewalls operating under
different policies \cite{ioa00-1}.
 Such diffusion tends to be complex, costly, and error prone.
However, no such diffusion is necessary in the case of FDS, for the simple reason
that the selection of laws to operate under is done by the actors
themselves, as has been discussed above.

\ss{The  Trust Modality Induced by this Architecture}\label{L-trust}
The effective operation of  a distributed system requires a degree of trust
between interacting actors.
Unfortunately  there is generally very
little basis for trust  between  the actors of an open distributed system.
The FDS architecture provides for a useful mode of such a trust---not between the
 actors themselves, but  between the $F$-agents animated by such actors.
This mode of trust is defined by the following properties of the FDS architecture:
\begin{enumerate}
\item One can trust the observable---i.e., interactive---behavior of every
      $F$-agent to comply with the law under which it operates.
This is the consequence of the fact that LGI-laws are strictly enforced.

\item  An $F$-agent $x$ can recognize if its interlocutor $y$ is also an $F$-agent,
and it can recognize the law under which $y$ operates, as well as the position
of this law in the the hierarchical structure of the framework. Moreover, $x$ is able to
get the text of the law of its interlocutor from the law-server. (These
properties have been introduced in \secRef{enforcement}.)

\item All  $F$-agents in a given FDS can be trusted to comply with the
irreversible provisions of the root law \law{R} of $F$, because they dominate
all other laws.
\end{enumerate}

\noindent
We call the the mode of trust resulting from these  properties
\emph{law-based trust}, or \emph{L-trust} for short.
This mode of trust, which is independent of the code of actors that animate the $F$-agents in question, is
fundamental to FDS as it facilitates  some   of its  basic features, such as
the easy of interoperation between $F$-agents. (for a more
comprehensive discussion of L-trust see \cite{min03-6}.)

 \ss{Interoperation Between $F$-Agents}\label{interop}
 Different $F$-agents operating under different laws in $F$ 
 often need to \emph{interoperate}, i.e., to interact with each other without
 violating their own laws.
 Although this is analogous to
 interoperability under conventional access control (AC), our interoperation
 mechanisms is different, and far simpler, and more flexible,  than that under AC.

The conventional AC approach to  interoperability  \cite{mcd02-1} 
 between parties operating under
 policies $P1$ and $P2$, respectively,
has been  to  \emph{compose} these policies into
a single policy $P12$, which is, in some  sense,
consistent with both $P1$ and $P2$. The composition $P12$ is then to be fed into an
appropriate reference monitor, which would mediate the interaction between the
two parties. Unfortunately,   composition of policies has
several  serious drawbacks: (a) manual composition is laborious, and error prone;
and (b)  automatic composition is computationally hard \cite{mcd02-1}, and
often impossible because the two given policies are inconsistent.
Yet, composition is the natural, and perhaps necessary,
 approach to interoperability
under  AC---because AC employs
a single reference monitor to mediate the interaction of any pair of agents.

 On the other hand, under FDS (and more generally, under LGI), composition of laws is neither
 natural nor necessary.
This is because of dual mediation for every pairwise interaction.
To see this, first
consider two actors $x1$ and $x2$, operating under laws 
 \law{1} and \law{2}, respectively. 
Due to the dual control over  interactions under LGI, via two separate
controllers (as shown in \figRef{fig-agent}), there is no need to compose 
\law{1} and \law{2} into a single law in order to enable $x1$ and $x2$ to
 interoperate.
Rather,  since  each of this laws  can recognize the other---due to
L-trust---they can specify their conditions, if any, for 
interoperation with it.

Moreover, 
 since the two interacting laws belong to a conformance hierarchy $F$,
it follows that they both conform to the   law \EL\ that is their lowest
common ancestor in $F$; in particular, all the law in $F$ conform to the root
law \law{R}.
 If this commonality between \law{1} and \law{2} is sufficient for them to
 interoperate then they can do it seamlessly.

\ss{Rogue Communication, and its Limited Effect on an FDS}\label{rogue}

 While the framework of an FDS has complete control over the interactive
activities of its $F$-agents---i.e., of the messages sent and received by
them---it 
does not,  generally,  control all the flow of messages in the system at hand.
The reason for this is that under most circumstances (described in paragraph
(2) below)  any
actor can engage in ``direct communication'' (via TCP/IP,
say), not subject to any law in $F$.
In the context of an FDS, we call such  communication \emph{rogue}, because it is not bound by the framework of this system.
  (\figRef{fig-arch}
depicts  such communication by 
the dotted arrow  from   actor $b$  of $S$ to the
external actor $u$---this message  is rogue in that it is a direct message, not
mediated by any controller, and thus not subject to $F$.)
It should be pointed out, however, that a framework $F$  may allow some
$F$-agents to communicate with certain actors not operating under $F$---particularly
with Internet cites that do not belong to the FDS in question.
But such messages are not rogue, as they are regulated by $F$.

 Obviously, rogue communication  can undermine the control that a framework has over an
 FDS.  Suppose, for example, that $F$ blocks all communication with a certain
 website $w$. This means, of course, that no $F$-agent   can communicate with
 $w$.  But the code of an actor that animate some $F$-agent can do so by
 direct messaging, and thus can reveal some information that should not be
 shared with $w$.
 This is, of course,  a general problem,  not specific to FDS or to LGI.
For example the access control imposed by the reference monitor of the XACML
mechanism~\cite{god05-1} over the components of an enterprise, does not really
determine who can access whom, because components can simply bypass this
reference monitor.

  Yet, as we shall see in the following paragraph, the
ability of rogue communication to undermine the provisions of the framework of
an FDS is limited, due, in part, to the existence of \emph{L-trust}.

\p{(1) The Limited Effect of Rogue Communication on an FDS:} Even if any 
actor can use rogue
communication, one can expect most functional communication  in an FDS  to be
done by $F$-agents. This is because actors  may be \emph{virtually compelled}
to operate as an $F$-agent,   if they need the services  of
 some $F$-agent, whose law does not allow communication with non $F$-agents.
 And often it is sufficient to know for a fact that 
 a single actor operates only as an $F$-agent, for the need to operate under
 $F$ to cascade through the
 system, establishing  a global, or semi-global, system property despite the possible
 presence of rogue  communication.
We demonstrate such virtual enforcement with the following example.

Suppose that our example-system Acme provides many disparate internal services,
which are to be used subject to
 following \emph{budgetary control} (BC) protocol, described informally below:

\begin{quote} 
\underline{\emph{The BC Protocol:}}
(a)   A distinguished actor called \emph{budget-office} has the exclusive
      role of providing---via  appropriate messages---system components with 
their \emph{service budget}, usable for any internal service orders;
(b) system components never overspend their service budget;  
and (c)    services  can  report to the budget-office their correct income for
services,  as accumulated
in  the state of their controllers---and we assume that services get credit for
such reported income.
\end{quote}
\noindent
If this protocol is  actually observed, everywhere in the system, it would have the following
consequences: 
(1) it would provide the budget-office with the ability to impose upper bound
    on the total cost of the service orders made by every system actor; and (2)
 it establishes a reliable means for services to report their correct income to the budget
office, thus getting the credit they are entitled to.
To ensure these consequences we can define this protocol
  as a law \law{BC}, and incorporated it in 
framework of Acme, in a manner to be  discussed in
\secRef{crosscutting}.

We now show that if we can assume that 
 the budget-office
interacts with servers and their clients only  subject to  law \law{BC},
then the above consequences would be virtually ensured, despite the ability of both
servers and clients to use rogue communication. This for the following sequence
of reasons. First, a   given service $s$ would have to receive service orders only under law
\law{BC}, otherwise it will not get its income recorded in a form that
can be sent to the budget-office, which accepts income reports only under law \law{BC}.
 This means that 
the clients of $s$ must send their service orders while operating under
law \law{BC}, for these orders to be received by $s$.
 This, in turn, implies that clients would not be  able to  exceed the budget provided to
  them by the budget-office.
And this, finally, means that the budget-office has control over the amounts of
``money'' that any given $F$-agent can spend on service orders.

So, law \law{BC} has the effect it is designed for, despite the ability of
both the servers and their clients to use rogue communication.

\p{2: Complete Elimination of Rogue Communication:}\label{antiRogue}
  Under certain circumstances   rogue communication can be blocked altogether.
This is the case if the system in question is confined  within an Intranet, or within  a set of Intranets managed under a
single administrative domain. Under these conditions
 one can force all actors in $A$---and the computers that host them---to communicate only as F-agents, by controlling the network, or
networks, in which these host operate.
For example, such control has  been  exercised for  the use of LGI to control
the usage of  distributed file systems
   \cite{ngu06-1}---via the firewalls attached to individual hosts.
A more systematic way for doing so should be possible under
\emph{Software–Defined Networking} (SDN) \cite{men12-1}.

\ss{The Self Regulatory Nature of the Framework of an FDS:}\label{self-reg}
Both the  base system of an FDS
 (i.e., its set of actors), and its framework, are bound to evolve.
 The evolution of the base system of an FDS presents no new difficulties. Quite the
 contrary, such evolution becomes safer under  FDS, because the 
 system properties establishes by the framework 
 are invariant of changes in the code. 
This is, indeed,
 one of the most significant advantages  of the concept of FDS. 
But changing  of the framework of an FDS  is far more problematic---not
surprisingly, because the framework controls the behavior of the system.
One of the problems involved with framework changes is discussed below, along
with its resolution. Other problems, mentioned briefly in \secRef{future},  are still open.

One of the issues involved with the changing of a framework  is that such a
change can have a
very powerful, and possibly harmful, effects on the system as a whole.
 These include the  disruption of some normal operations of the system; 
and  the lowering of  the defenses  against  attacks,  thus compromising its
 security and dependability.
It is therefore critically important to 
 avoid careless or
malicious framework changes, by \emph{regulating its
process of evolution}. In other words, it should be
possible to control who can carry out which kind of framework changes, 
  and under which circumstances should such changes
 be permitted.  Moreover, if changes of a framework are
to be made by different stakeholders, it may be necessary to establish 
coordination protocols between them.

Such regulation can be readily accomplished because the framework of an FDS is
naturally \emph{self regulatory}, in the following sense.
 The framework $F$ of a given FDS is maintained in the law-server $LS$, which is,
itself  an actor of the system in question. 
Therefore, since changes of $F$ must be carried out by means messages sent to $LS$,
and since these messages are governed by $F$, it follows that  $F$
can regulate  its own evolution.
This is a  clean, and potentially very important, property of the
FDS architecture.

\s{An Implemented  Case Study---a Summary}\label{case}
As a  concrete view of an FDS, and particularly of  its framework,
 we provide here a simplified outline of a case
study  we have implemented. Our intention is 
 to illustrate the nature of a  framework, and some of the impact
it can have on the system governed by it. The case study uses
 the Acme enterprise system introduced in \secRef{intro}, assuming now that this system  consists of  two  divisions,  $D1$ and $D2$, 
serving  two semi-independent branches, which are
 implemented and maintained under different administrative  domains.
 Acme also contains  a set of actors
that serve the enterprise as a whole, which includes, among others:
 (a)  a certification authority (called \emph{AcmeCA})  that
 provides each actor with a digital certificate that identifies it; and 
(b) the law-server ($LS$) that maintains the law ensemble that constitutes the
    framework  of the system.

 The framework $F$ we envision for this system is a three level
conformance hierarchy of laws, depicted in
\figRef{fig-laws}. 
The first level of this hierarchy---introduced in \secRef{global}---consists
of the root law \law{R}.
The second level contains the laws of the two divisions of the system,
introduced in \secRef{divisions};  as well as  law \law{BC} that crosscuts through
both division,  and is discussed  \secRef{crosscutting}. 
The third level, introduced in \secRef{actorLaws}, is a
collection of laws that governs the interactive activities  of some individual
actors. 

All these laws  are described here informally, and they reflect only part of
the laws used in our case study.
Note  that all but the root law
 are represented by their deltas\footnote{The concepts of  \emph{delta}, and of
the \emph{meta} part of a law (mentioned below) have been introduced
in \secRef{conformance}.}, which specify the
 differences between the law at hand and its superior law.
A reader who wishes to see how such laws are actually written  is
referred to \cite{min03-6}, where a fairly sophisticated hierarchical
ensemble of laws is introduced in details. Here are some comments about our
informal description of  $F$.

In our informal description of these laws   we employ  the following
convention about the meta part of any given law \CAL{L} in the hierarchy: (a)
if \CAL{L} has a rule that addresses a certain aspect of interactive activity
of an actor subject to it---such as the sending of a certain type of
messages---then this rule is \emph{irreversible}, i.e., it cannot be deviated from by subordinate
laws of \CAL{L}, unless such deviation is explicitly permitted by the meta part
of \CAL{L} (such meta permissions are denoted by bracketed texts in bold
italics);
 and (b) if \CAL{L} is silent about certain aspect of
interaction, then subordinate laws have the freedom of legislation about it.

\ss{The Root Law}\label{global} 
As has already been pointed out, the root law \law{R} is the global law of the system, in the sense that
all its provisions are shared by all the laws in $F$---modulo 
 modification by
 subordinate laws, if permitted by \law{R}.
 The main role of \law{R} is to establish broad system
regularities and defaults.
 The following is the set of  rules that govern Acme, which should be viewed as
 a small sample of rules that can be established by such a law.
 We elaborate on these  rules in the discussion that  follows them.

\begin{enumerate}
\item  \textbf{Authentication of actors:}
To  adopt an LGI-controller under this law,
an actor $x$ needs to authenticate itself
 via a certificate signed by $AcmeCA$, which we assume to  identify the unique name of $x$, 
with respect to  the Acme system, and
 the division to which it  belongs.
This authenticated identification of 
 actors is  stored in the state of their
adopted  controllers.
\textbf{\emph{[Subordinate laws may add conditions to this
rule, and may require additional  operations to be carried out upon adoption,
but they cannot weaken this rule.]}}
\label{R-auth}

\item
\textbf{Sender identification:}
 Every message sent is to be concatenated with the name and division of its
 sender.  This identifying  information would be visible to the
 controller of the recipient of this message, but would be
stripped from the message before  it is delivered to the target
actor---although the actor can get it upon request.
\label{R-id}

\item \textbf{Constraints  over the interaction between $F$-agents:} The following
      two
 provisions are made by this
                rule:   (a) all inter-division interactions are prohibited,  \textbf{\emph{[unless
 permitted by the corresponding subordinate division laws]}}; and (b) all intra-division interactions
 are permitted, \textbf{\emph{[unless prohibited by the subordinate division
 law in question]}};
\label{R-AC}

\item \textbf{Establishing an Audit trail of inter-division interactions:}  Every inter-division
      message would be logged
      in a specified logging service, upon its arrival.
\label{R-audit}
   
\item \textbf{Providing a manager with the power to   control:} An actor of
      the base system that receives a message  \texttt{stop(pattern)} sent by
a  distinguished  $mgr$    actor, would  lose
 the ability to send or receive messages of the specified pattern, which may be
 ``all''.
\label{R-mgr}

\item  \textbf{Rate control:} The rate control protocol (RC) introduced in
       \secRef{intro}  is part of \law{R}---in fact, the protocol actually
       established by $R$ contains the generalization of RC alluded to in  footnote~\ref{foot-a}.
\label{R-RC}
\end{enumerate}

\p{Discussion:}
The following is  an elaboration on these rules, which provides
some  clarification and motivation for them.

\ruleRef{R-auth} provides Acme  with a degree of
control---exercised via its CA---over which  actors can  operate as $F$-agents,
and under which laws in $F$.
Note that this provision is irreversible, governing all $F$-agents, although it can be tightened by
subordinate laws.
 Also, note that maintaining the certified identification of each actor
 in the state of its  controller
 facilitate the enforcement of other 
 rules of this law, such as rules~2~and~3.

\ruleRef{R-id} provides the receiver of a message the ability   to identify its
sender. This is more informative and more   trustworthy than 
identifying the sender 
 by its IP address, which may not carry much meaning to the receiver, and which
 can be spoofed.
Such identification can be useful in many ways. In particular it is used here
 for enforcing the constraints of \ruleRef{R-AC}.

\ruleRef{R-AC} establishes two different types  of access control provisions:
 Provision (a) prohibits all  inter-division interaction, as a default, 
allowing subordinate  division laws  to permit any such interactions (see
\secRef{divisions} for how this can be done).
(Note the  unconventional nature of this type of conformance, where the
subordinate division laws can be more permissive than their superior law.)
 Provision (b) is analogous to (a), with the opposite effect.

\ruleRef{R-audit} ensures dependable logging  of  all
inter-division messages. Note that 
 this rule can be stated here  despite the fact that \ruleRef{R-AC} of the
same law prohibits all inter-division messages---but if such an interaction would be permitted by the subordinate division laws,
it would be subject to
this rule.

 \ruleRef{R-mgr}  enables system managers operating via the
distinguished actor $mgr$ to prohibit any given actor from sending and
receiving messages that fit a specified pattern.
Such prohibition with the patter ``all'' would effectively remove the actor in
question
 from a system by stopping all its
communication.
This is just an  example of how one can endow system manager with a real
power over the distributed system  it is managing.

Finally, \ruleRef{R-RC}, which  implements the  rate-control protocol introduced in
\secRef{intro}, must be part of the root law, as it apply to the entire system.

\ss{Division Laws}\label{divisions}
A division law, say  law \CAL{L}$_{D1}$ of division $D1$,  is to be derived from
the root law \law{R} via a delta $\Delta($\law{R},\law{D}$_1)$.
One can reasonably assume that the writer of this delta has  some idea of the
intended structure of this division, and on the intended role and function of
certain of its actors. This delta can, then, be used to impose this
structure.
For example, the delta of   law \CAL{L}$_{D1}$ may make the following three
types of provisions.

\p{(1)  Constraint on the Composition of $D1$:}
Given that  Rule~1 of law \law{R} permits its  subordinate laws
to add conditions on their adoption , this delta may require that actors
adopting law  \CAL{L}$_{D1}$  would be authenticated as belonging to division
$D1$.

\p{(2) Imposing Constraints over Intra-Division Interaction:}
Recall that all intra-division interaction have been permitted  by \law{R},
as a default, allowing subordinates laws to impose arbitrary prohibitions on such
interactions.
So, this delta can impose any desired prohibition on the interactions between
 $F$-agents  belonging to $D1$.

\p{(3) Enabling Selected Inter-Division Interactions:}
Recall that  inter-division interactions 
are  prohibited by law
\law{R}, as a default,  allowing subordinates laws to permit them.
Note,  however, that  for an interaction between $D1$ and $D2$ to  be enabled,
it must be permitted by both  \CAL{L}$_{D1}$ and  \CAL{L}$_{D2}$---this is  due to the local
nature of our laws.
 For example, to permit a message from  an $F$-agent $a1$ in $D1$ to an $F$-agent
 $a2$ in $D2$, law \CAL{L}$_{D1}$ needs to permit $a1$ to send a
message to $a2$, and  \CAL{L}$_{D2}$ needs to permit $a2$ to receive
this message.  Of course, such a permission may be formulated to apply to whole
sets of interaction types; for instance, the laws of the two divisions can have
rules resulting in enabling  certain types of
messages to be exchanged between a certain pairs of $F$-agents belonging to the
two divisions.

\ss{Laws of Individual actors:}\label{actorLaws}
An actor $x$ belonging to a certain division, 
say $D1$, may operate directly under law \law{D}$_1$. 
But $x$ may chooses to 
 operate under its own law \law{x}---subordinate to \law{D}$_1$. One reason for
 $x$ to do so can be  as follows:
Suppose  that $x$ is a web server,   and that it makes certain promises
to its clients about the services it provides. 
But such promises are not very credible if they are just stated, on the
website of $x$ say------particularly not in an open system, where
the code of the service is not known to its clients, and where this code can be
changed without the client's knowledge.
This is a serious and well known difficulty with services over the Internet.

 However,
promises that can be formulated in terms of message exchange can be rendered 
trustworthy and dependable by formulating them as a law, 
 and then  providing one's services via a controller that enforces this law.
The clients of such a service can trust these law-based  promises
  due to the existence of L-trust, which has the following consequences: (a) the promise can be verified by studying the
law---which is likely to be
  much smaller and simpler than the server's code; and (b)
 the law cannot be changes without the client's knowledge.

There are many examples of important promises that can be  
rendered trustworthy in this way, including such things as \emph{money back
guarantees}, the so called \emph{service level agreements} (SLAs),
\emph{confidentiality}, etc. And, as pointed out before, a single actor
may form different $F$-agents operating under laws that make different types of
such promises. Below is a more detailed, but still informal,
discussion of one type of such promises.

\p{Server's Promise Made  During a Conversation:}
The interaction between a server $s$ and its clients may involve a
sequence of messages exchanged in a predefined order, called
\emph{conversation} \cite{casati-soa-03}. During  such a conversation,
the server may make various promises to the client. For such promises to be dependable,
they need to be enforced.
For example, suppose that our server is  a travel agent that provides for the
following kind of conversation:
 A client $c$ may request
to reserve the right to buy a certain ticket at a particular price $p$, within
a  grace period $t$. If the server agrees, it should sell that ticket
to $c$, if $c$ pays for it within period $t$---which means that
the server should not sell that ticket to anybody else within
this time period.
Of course,  a law  \law{s} that formalizes such a promise \emph{must conform to its superior law}. 
 (Note that an LGI law that  enforces such a 
promise, in a different context,  has been described  in  \cite{min07-2}.)

\p{The Global Effect of of Local Laws of Individual Actors:}
It is worth pointing out that although the law \law{x} of an actor $x$ effects
directly only the interactive activities of $x$ itself, it has a global effect on
the system in that it engenders a degree of justifiable trust in the behavior of
$x$,  by every actors in the system.

 Moreover, having all, or most, actors in a
distributed system define their promises in this manner, can have an important
 effect on the dependability of a system at large, as argued by
Burgess \cite{bur05-1}.

\ss{Crosscutting Laws}\label{crosscutting} 

Consider a group $G$ of actors  dispersed throughout the
system, so that some of its members belong to division D1, while other belong
to D2. And suppose that members of this group need to interact with each other
subject to a law \law{C}---''C'' for crosscutting---such as law \law{BC}
 introduced in \secRef{rogue}.  The question is, how do we incorporate 
 law \law{C} into the framework
$F$ of the system, so that all members of group $G$ can operate subject to it.

Using law  \law{BC} as an example, it cannot be defined
as  subordinate to either  law \law{D1} or  law \law{D2}
because group $G$ crosscuts through  them.
And we  cannot define \law{BC} as 
part of the root law \law{R} of $F$, because \law{BC} is not supposed to govern the entire system.
So, we place this law as subordinate to \law{R}, as is shown in
\figRef{fig-laws},
and we provide two different ways for the members of $G$ to operate under it.

The simplest way for members of group $G$ to operate under any given law  \law{C}, which is
subordinate to the root law \law{R}, is based on the 
 ability of any  actors
 to operate, simultaneously, under several different laws---provided that these
 laws agree to be adopted by the actor in question.
But this technique suffers from the following limitation: Consider an actor $x$
that is designed to operate only under a given law \law{x} (which may be its
own law, as discussed in \secRef{actorLaws}, or one of the division laws.) 
If actor $x$ would operate under another law \law{C}, it would not be
subject to the constraints imposed by \law{x}---which may be undesirable.

\begin{figure}
\leavevmode
\epsfysize=1.5 in
\epsfxsize=6.0 in
\centerline{\epsffile{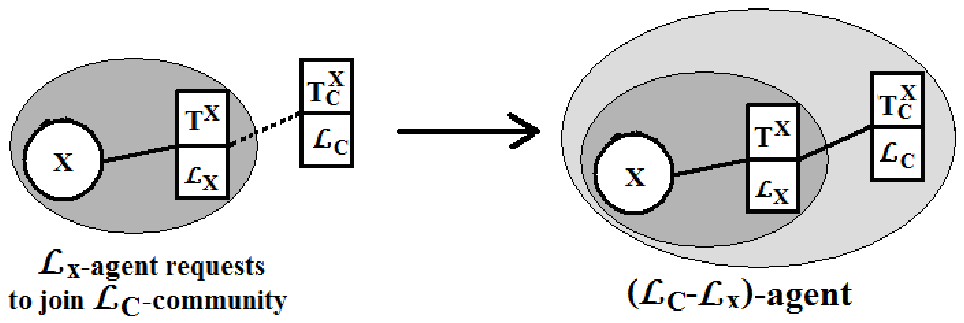}}
\caption{Operating Under Two Laws}
\label{crosscut}
\end{figure}

We have devised, therefore, another way for 
 enabling every member $x$  of a crosscutting group $G$ to operate under a common
law \law{C},  while  also operating subject to its native
law \law{x}.
Note that so far LGI allowed only a bare actor $x$ to operate under a some
law \law{x} thus forming what we call an \law{x}-agent, which is a pair
$\langle x,$ \Tla{L}{x}$\rangle$.  We now a allow an \law{x}-agent to adopt
another law, say \law{C}, which would mean that actor $x$ operates under a
 series  of two  controllers, subject to two laws, \law{x} and \law{C}, correspondingly---as
 depicted in \figRef{crosscut}. And it should be pointed out that the adoption
 of law \law{C} by    an \law{x}-agent, is subject to approval by both
 laws, \law{C} and \law{x}. For more detailed description of this feature of
 LGI see \cite{min12-1}.

It is  worth pointing out, that this facility constitute an inherently
distributed treatment of the \emph{crosscutting concerns}, the underlying 
 concept of \emph{aspect oriented programming} AOP
\cite{kic05-1}.

\s{On the Impact of the FDS Architecture on the Dependability of Distributed Systems}\label{impact} 
We identify here three of the   modes in which the FDS architecture can impact the
dependability of distributed systems. All of them involve \emph{system
properties}, which can be established via the framework of the system in
question. These modes of dependability are the following:

\begin{enumerate}
\item \emph{Ease of  implementation and verification:}  It is easier to
      implement a given   system property $P$, and to verify  that it is satisfied by the system, when  it is established via a framework $F$, than 
 when it has to be implemented correctly in many parts of the system.

\item  \emph{Independence of the code:} A property $P$ established by the
       framework of an FDS is inherently  independent of the code of the system, and thus invariant of changes in
      it. So, once $P$ is verified it can be dependent on to be satisfied while
      the code of the system evolves.

\item  \emph{Stability:} While a  property  established by the framework $F$
       can be changed by changing $F$,  changes of a framework  can be
       regulated (cf. \secRef{self-reg}), which can  enhance the stability of
       the properties defined by it. 
\end{enumerate}
\noindent
These  mode of dependability have fairly  broad implications, as they can be applied to a wide range
of system properties, involved in various aspects of a system.
 We limit our discussion here to  
 the potential impact of FDS on the \emph{fault-tolerance} (FT) of distributed
 systems, focusing on the fault-tolerance at
the \emph{application level of  heterogeneous and open systems},
which, as we shall see, is  hard to achieve via the conventional means.

  The need to develop fault tolerance techniques specifically for the
  application level of systems---sometimes called ``software
  fault-tolerance''---has 
 been pointed out already in 1975 by Randell \cite{ran75-1},
who argued that the traditional FT techniques, designed mostly for hardware
failures, are not sufficient for handling the various ways in which an
application 
may fails. This is
 true, in particular, for  \emph{coordination failures}.
Such as  a failure of 
a group  of distributed  actors  to collaborate effectively towards a common
goal, or to compete safely over some resources, 
 due to the failure of  any
one of them to abide by the necessary coordination protocol. (In an analogy,  consider
 what may  happen when one car in an intersection does not stop on a red light.)

 Considerable  research effort has been devoted to application
level FT since Randell's paper---see \cite{flo08-1} for a survey.  This
generally involves incorporating some failure-handling code 
into the software.
 Various types of FT measures have been developed in this way, for dealing with
 various situations---such as
 \emph{exception-handling}, \emph{recovery
blocks} \cite{ran75-1},  \emph{N-version programming} \cite{flo08-1}, and 
 \emph{coordinated atomic actions} (CAA) \cite{ran95-1}. The deployment of
such techniques suffer from two types of difficulties, even in non-distributed systems: (a) they tend to complicate the
system, and (b) when the FT-measure in question require the incorporation of
corrective code in many system components,
their deployment tends to be laborious and error prone, even if the same code
needs to be thus incorporated. These  difficulties can
sometimes be alleviated 
via  \emph{meta object protocol}
(MOP) \cite{kic91-1}, which enables what is called \emph{reflection}; 
or  via \emph{aspect oriented
programming} (AOP) \cite{kic01-1,zar06-1,cap07-1}.
Moreover, special programming languages, such as
Argus \cite{lis88-1}, and coherent sets of tools, such as Arjuna \cite{shr95-1},
were developed  for building fault tolerant  systems.
Some of these techniques can be applied even for distributed systems, if they
are  monolithic. That is, if a system is 
designed and maintained under a single administrative domain, and if it
employs a single language.
This is the case, for example,
when one can ensure that all system component are governed by the same AOP
code.

 But such
code-based FT measures are generally unsuitable for open systems, due to the lack
of overall control over the code of the various components, or even of the
language in which they are written.  \emph{This leaves open distributed systems 
vulnerable to their own faults, and to attack on them.}

However, a  substantial range of  FT measures can be
established  by controlling the flow of messages in the system, independently of the code
 of the communicating actors.
 Of course, this cannot be done for  all FT measures that can be
 established---in monolithic systems---by 
 inserting suitable code into the
components themselves.
For example,  controlling messages cannot ensure orderly checkpointing by
selected components---an important basis for many conventional FT-measures.
Yet, as we  demonstrate below, there is a substantial range
of FT measures that can be established via the framework of an FDS,  either
completely  by controlling messaging, or  with the help
of relatively few distinguished actors that can be trusted to carry out the
role assigned to them.

Moreover, although such framework-based FT-measures are necessary for open systems,  some of them can be useful for distributed systems in
general, even were traditional code-based techniques are feasible.
This for two main reasons:  first,  our  FT-measures would
be independent of most of the system code, and
cannot be violated by changes in it---a distinct advantage in any system.
 Second, enacting such measures would not complicate the code
because  the framework is completely separate from it---this is, in a sense, similar to FT measures
implemented via the
meta-object protocol, or via AOP, which we can independently of the language in
which the components are written.

\ss{Application Level Fault-Tolerance Under an FDS}
We consider below examples of framework-based measures that span the
following  aspects of fault tolerance: (a) preventing failures; 
(b) isolating a system from misbehaving actors;
(c) recovery from failures;
and (d) reconfiguration.

\p{Preventing  Failures:}
 Failures can sometimes  be prevented  by imposing a structure on a given system
 that helps  in avoiding situations that may lead to certain types of
 failures;  or 
by providing means for averting failures by actively stopping behavior which
would cause a failure if allowed to continue.
Below are examples of these two types of prevention.

\noindent \emph{(1) Preventing Coordination failures:}
Consider a group
$G$ of distributed actors that need to coordinate their activities, subject to
a given protocol $P$. As already pointed out, such
 coordination may fail due to
any member of $G$ not following protocol $P$.   There are plenty of
 general purpose protocol of this kind, such as  the \emph{token-ring} protocol
 for ensuring   mutual exclusion, and  protocols   for
  \emph{leader election}. And there  are many types of  potential
 application-specific protocols, such as  the  protocol 
introduced in \cite{wey10-1} for a  collection of
cameras that monitor road traffic.
Generally, one assumes that all participant in a given coordination activity
abide by the protocol designed for it. But this assumption are mostly
unwarranted in open systems.
This is notably the case for
\emph{choreography}  \cite{par09-1}, namely the interactive coordination between web-services.
Although a language for describing  choreographies have been devised by W3C, it
is not  executable,  and certainly not enforceable.

Under FDS, however, if a  protocol $P$ can be formulated in terms of message
passing, then it can be expressed via a law \law{P} of the framework, which is to be employed by
all members of group $G$ for their coordination activity---this is possible
because LGI-laws are sensitive to the history of interaction, and because they
are proactive---that is, they can force some actions to be carried out, via a
mechanism of enforced \emph{obligation},
which can ensure  a degree of \emph{liveness}. A case in point, implemented
under Acme, is the $BC$
protocol described in \secRef{rogue}.

\noindent \emph{(2) Preventing Denial  of Service:}
The    \emph{rate
control} protocol RC defined in the Introduction, and 
 implemented over the Acme case study,
 enables a distinguished
server $v$ to protect itself from \emph{denial of service}, essentially by forcing
any other actor in the system to obey the \texttt{slowDownTo(r)}  messages sent
to it 
by $v$. (This is an  application-level version of the flow control of
TCP/IP.)

\p{Isolating a System from Misbehaving Actors:} 
The framework $F$ of an FDS can specify the type of messages that any given
actor $x$ can send.  And if an actor $x$ misbehaves by sending a message it is
not entitled to---a message that may cause some damage to its target---this message can be blocked by the framework, thus protecting the
system from being effected by such a  rogue actor. Moreover, the offending
message may be logged, thus enabling the identification of the rogue sender,
and perhaps its eventual removal.

\p{Recovery from failures, or Self-Healing:}     
For a recovery mechanism to be effective it should be able to handle a
reasonably wide range $R$ of failures, by a possibly heterogeneous group
$G$ of actors. For example, the range of failure in question may consist of 
inappropriate sending  of  purchase orders (POs) to outside
vendors, where purchase orders may be inappropriate in many different ways.  And the set $G$ of actors that may be implicated in such failures, may
consist of all system actors, most of which may have no right to send POs, but
which can attempt to do so nevertheless.

 To carry out such a recovery mechanism, a  prospective \emph{healer} $H$ needs
to be able to: (1) sense all the activities of all members of $G$,
 which are relevant to $R$; and (2)
 exert a degree of control   over the failing actors in $G$, in order to heal
it, or to protect the system from it.
But since in a distributed system nobody has an intrinsic ability
to either sense or control other actors, the above capabilities require 
certain \emph{regularities} in the behavior of actors in $G$,
despite their possible heterogeneity.
In particular, all actors in $G$ need to send to the would be healer $H$ copies
of the messages it sends and receives, which may be relevant for discovering
failures in $R$. And all these actors should obey certain types of commands
sent to
 them by $H$,
such as the command to stop sending certain, or all, messages.
And  such regularities must, of course, be  invariant of the failures
in range $R$
of the actors in $G$.

Establishing such regularities by the individual actors themselves,  via individual
wrappers \cite{voa98-1}, say, is laborious and error prone even for monolithic
systems, and it is next to impossible to do so reliably in open systems.
But it is often possible to  do so by controlling the flow of messages, as we have
demonstrated in \cite{min03-5}; and can, thus, be done via the framework of an FDS.

\p{Reconfiguration:}
Recovery from failures often involves reconfiguration of a system.
 Most current approaches to reconfiguration
(see \cite{zha05-1}, for example) employ
central manager, which is assumed to 
have sufficient  knowledge of the system to carry its task. But there is a
growing realization that reconfiguration often require coordination between
distributed actors \cite{arb01-1,zar06-1,wey10-1,ran95-1}, rather than being managed
centrally. 
And  we content that such reconfigurations can often  be facilitated, or
completely accomplished, by means of the framework of an FDS

A  case in point is a  set of actors  engaged in a \emph{token-ring
protocol}, where  the ring needs to be reconfigured dynamically by removing failing actors from
the ring, and by adding new actors to
it---without having to stop the operation of the ring, and without loosing or
duplicating the circulating token. This can be done when all actors involved  in such    reconfiguration
comply with  a suitable  reconfiguration protocol; but in an open systems one cannot generally
rely on such compliance, or indeed---on the compliance with the basic token-ring
protocol itself. Having this problem in mind we have designed  \cite{min95-8} a token-ring
protocol which lends itself to a safe reconfiguration  and we
wrote an
 LGI-law that establish this protocol. This law can be easily incorporated
into the framework of an FDS.

\s{Related Work}\label{related}
Work related to  fault
tolerance are discussed  in 
\secRef{impact}. Here we  discuss only  work related to the concept of FDS.

The literature is replete with papers that identify their subject matter by
phrases such as ``policy based
frameworks,''  ``policy based systems,'' etc.
But we  focus here only on papers that share our  objective to  be applied to
 open distributed  systems. This means, as pointed out by Principle~1 in \secRef{principles},
 that the framework can only control the flow of messages between
system components, while being  oblivious of their  internals.

These excludes many papers that provide systems with some kind of framework. Such as 
 papers about ``software architecture''  \cite{gar95-1}, which is an unenforced  specification of a
system. We also exclude papers that make strong assumptions
about the code of system components. Such as paper that use
 ``aspect oriented
programming'' \cite{kic05-1}.  And like
  \cite{row04-1} and
\cite{cap07-1} 
which are Java-based.
Still another class of papers that we exclude from  consideration here,
 follow the IBM approach to 
\emph{autonomic system}  \cite{whi04-1}, which expect
each system component to be \emph{autonomic}. 
Similar assumption are made by  \cite{per10-2}, whose 
 framework is meant to  provide dependability management.

Examples of paper that share are main goal---to which we  refer  below as IBF
papers
 (IBF for ``Interaction Based
Framework'')---includes the following, among others:
\cite{bel02-1,lee11-1,li11-1,ioa00-1,rib07-1} (but note  that some of these
papers
  do not use the term ``framework'' explicitly.)  These papers control the flow
  of messages in a system, independently of the code of the interacting components,
mostly  via  some kind of access control (AC) mechanism, such
as XACML~\cite{god05-1} or RBAC~\cite{osb00-1}. 

However,  by and large, none of these papers supports most of the principles of FDS
introduced  in  \secRef{principles}. We  review below some of the
IBF mechanisms listed above in the context of some of these principles.

\begin{itemize}
\item \emph{Scalable sensitivity to the history of interaction (Principle~2):}
      Only one of
IBF mechanism \cite{rib07-1} is sensitive to the history of interaction, but it
is far from being scalable, as shown in \cite{min12-2}.

\item \emph{Trust in the   interactive behavior of $F$-agents (Principle~5:)}
Such trust is realized under FDS by the concept of  L-trust introduced in
\secRef{L-trust}, and it is critical to the effectiveness of the FDS
architecture.
For example, this mode of trust facilitates the flexible interoperability under
FDS.

The closest that the IBF papers come to L-trust is trusting that no system
components would be able to violate the AC policy in question. But as we have
already pointed out, AC policies do not regulate the dynamic interactive
behavior of
 the various system components---at least, they cannot do so
 scalably. Therefore,  IBF papers do not support anything like our L-trust.

\item \emph{Decentralized enforcement (Principle~3):}
 most IBF mechanisms enforce their constrains over messaging in a
centralized manner.
A rare exception is
the use of distributed firewalls for regulating the flow of messages in an
 enterprise  \cite{ioa00-1}. But no scheme of distributed firewall know to us
 has anything like our conformance hierarchy for its policies---although this
 would have been very useful in this context. Moreover, this technique requires complex diffusion
 of the local policies that the various firewalls are to enforce, while under
 FDS no such diffusion is required, as explained in  \secRef{deploy}.

\item \emph{Modularity of the framework (Principle~6):}
Several IBF papers \cite{bel02-1, lee11-1}
 employ  concepts related to our conformance hierarchy---which is the basis for
 the modularity of our framework. But they differ from
 ours in several  ways. The most important of which is that
the conformance in all IBF paper known to us is not inherent to
the structure of the hierarchy, but needs to be verified, which generally needs
to be done manually.

\item \emph{The need for the framework of an FDS to be controllable
      (Principle~7):}
We know of no IBF project that attempt to have its framework controllable, much
less self-regulatory.
\end{itemize}

\s{Open Problems Raised by the FDS Architecture}\label{future}
The  concept of FDS,  as presented in this paper, raises some open  issues that
need to be  addressed for this architecture to attain its full potential. Two
of these issues are described briefly below.

\p{Converting  a Legacy System into an  FDS:}
Consider  a legacy distributed system consisting of a set of actors  $A_0$,
which is to be converted into an FDS,  subject to  
an initial framework  $F_0$.  Such conversion is critical for FDS to be widely
accepted by the industry. The main problem with such conversion
  is that  there are  bound to be conflicts between the framework $F_0$ and
 the actual behavior of the legacy systems $A_0$---conflicts that may disrupt the
 operations of the legacy system once  $F_0$  is imposed on it.
One needs to  develop tools for   identifying these conflicts, and for
resolving them by changing either $F_0$ or $A_0$, or both.
And since the conversion is bound to be a process rather than a one shot
affair, it would be necessary to develop   a methodology for carrying out the
conversion process.

\p{Issues Concerning the Evolution of the Framework of an FDS:}
We have already provided for the \emph{control} of the evolution of the framework of an FDS
(cf. \secRef{self-reg}). But \emph{carrying out} any change of the framework of a system presents some
technical difficulties, including the following:
(a) how to discover the  possible disruptive effects of a planned framework change, before
enacting it; 
(b) how to carry out a change in a non-leaf law, given the dependencies of
    subordinate laws on it; and   (c) how to carry out framework  changes
while the system continues to operate---this problem has been solved
  \cite{min09-2} for a system operating under a single law,  but doing so under
  a multi-law framework is much more challenging.

\s{Conclusion}\label{conclusion}
We have introduced in this paper a novel architecture  of distributed
systems---called  \emph{framed distributed system}, or FDS---that braces a given
system via a \emph{framework} that controls the flow of messages 
in it, while being oblivious of the code of the components that send and receives these
messages. The framework of an FDS is a highly modular collection of laws, which
are strictly enforced in a decentralized, and thus highly scalable manner.
Since the framework is enforced, in can be considered  as an integral part of
the system and not just an external specification of it.

While being applicable to any distributed system, the FDS architecture should be particularly
useful for for highly heterogeneous and \emph{open} systems.
This paper demonstrate the impact of FDS on the dependability of distributed
systems, focusing on the fault tolerance at the application level of such
systems.
Although  we expect
the FDS architecture to have a broad impact also on the security of distributed
systems, and on their entire life cycle, the analysis of such impact is beyond
the scope of this paper.

 It should be pointed out, that  this is a \emph{work in progress}, in two
respects.
First, although  the implemented 
case study of FDS, described in \secRef{case}, constitutes a \emph{proof of
concept} of this architecture,
 the  real usefulness and effectiveness of this architecture
 needs to be validated by applying it
 to   one or more real (or realistic) large scale and complex
distributed systems. Such validation is yet to be done.

Second, as stated in \secRef{future}, the FDS architecture introduced in this paper raises some open issues
that need to be addresses, for FDS to  attain its full potential.

\bibliography{../../../writing-tools/biblio,bills}

\end{document}